# Unveiling the Boson Peaks in Amorphous Phase-Change Materials

Jens Moesgaard[1], Tomoki Fujita[1], and Shuai Wei[1,2*]

[1]*Department of Chemistry, Aarhus University, 8000 Aarhus, Denmark.*

[2]*iMAT Centre for Integrated Materials Research, Aarhus University, 8000 Aarhus, Denmark.*

**Abstract**

The Boson peak is a universal phenomenon in amorphous solids. It can be observed as an anomalous contribution to the low-temperature heat capacity over the Debye model. Amorphous phase-change materials (PCMs) such as Ge-Sb-Te are a family of poor glass formers with fast crystallization kinetics, being of interest for phase-change memory applications. So far, whether Boson peaks exist in PCMs is unknown and, if they do, their relevance to PCM properties is unclear. Here, we investigate the thermodynamic properties of the pseudo-binary compositions on the tie-line between $Ge_{15}Te_{85}$ and $Ge_{15}Sb_{85}$ from a few Kelvins to the liquidus temperatures. Our results demonstrate the evidence of the pronounced Boson peaks in heat capacity below 10 K in the amorphous phase of all compositions. By fitting the data using the Debye model combined with the Einstein model, we can extract the characteristic parameters of the Boson peaks and attribute their origin to the excess vibrational modes of dynamic defects in the amorphous solids. We find that these parameters correlate almost linearly with the Sb-content of the alloys, despite the nonmonotonic behaviors in glass forming abilities. A larger contribution of excess vibrational modes correlates with a larger width of enthalpy relaxation below the glass transition temperature $T_g$. In a broader context, we show that the correlations of the characteristic parameters of the Boson peaks with $T_g$ and kinetic fragility, vary according to the type of bonding. Specifically, metallic glasses and conventional covalent glasses exhibit distinct patterns of dependence, whereas PCMs manifest characteristics that lie in between. A deeper understanding of the Boson peaks in PCMs holds the promise to enable predictions of material properties at higher temperatures based on features observed in low-temperature heat capacity.

[*] Corresponding author: shuai.wei@chem.au.dk (SW).

# 1. Introduction

In glasses, the heat capacity $C_p$ below 30 K shows distinct behaviors from that of crystals, where an anomalous hump in $C_p$ over Debye's prediction ($C_D \propto T^3$) has been ubiquitously observed[1–3]. The hump, known as the Boson peak, corresponds to an excess in the vibrational modes in the terahertz range, and is widely regarded as being universal in low-temperature amorphous solids[1,4]. The excess vibrational mode also underlies the anomalous plateau in the temperature-dependent thermal conductivity in glasses at low temperatures[5]. Although the Boson peaks are believed to be related to the structural disorder in glass systems, their exact origin has been understood in terms of various theories such as elastic heterogeneity[6,7], soft mode vibrations, and anharmonicity[8]. Recent computational studies have proposed that the string-like dynamic defects in amorphous solids, whose quasi-localized vibrations at low temperatures contribute to excess modes, are the origin of the Boson peaks[1,9]. Those defects provide also a common structural origin of the fast (β-) relaxations and slow (α-) relaxations in model glass systems[1].

The characteristics of Boson peaks have been shown to correlate with important material properties of metallic, oxide, organic, and molecular glasses [5], including kinetic fragility[2,10], glass forming ability[11], structural relaxations[1], shear modulus[12,13], and Poisson's ratio (i.e. mechanical properties)[14]. The fragility concept, proposed by Angell, describes the extent, to which temperature dependence of viscosity $\eta$ (or relaxation time) deviates from the Arrhenius law[15]. The fragility index $m$ is defined as $m = dlog\,\eta/d(T_g/T)|_{T=T_g}$, where $T_g$ is the standard glass transition temperature measured at 20 K min$^{-1}$. Some liquids such as silica ($m \approx 20$), showing a near-Arrhenius behavior, are classified as strong; others (with larger $m$ values), exhibiting a range of non-Arrhenius behaviors, are classified as fragile liquids[15]. Fragility is associated with spatial dynamic heterogeneities in glasses[16–18], and is related to the kinetics of nucleation and growth of crystals[17,19]. An earlier study showed that the Boson peak height decreases with increasing fragility in oxide, organic, and molecular glasses[10]. In metallic glasses, the excess heat capacity $C_p/T^3$ of fragile systems shows barely a hump, while strong glasses show larger humps with a lower maximum temperature[3]. A possible correlation of the Boson peak height with glass forming ability has been reported[11], although the validity of the correlation has not been confirmed in a wider range of systems. Systematic studies of Boson peaks in annealed glasses demonstrated that a decreasing height of Boson peak can be related to a larger enthalpy relaxation upon glass aging[20] and correlated with a smaller linear thermal expansion[21].

These correlations have been usually attributed to the aging-induced annihilation of elastic flow units[22], soft regions[23,24], or defects-like regimes[1] spatially distributed in the heterogenous structure of glasses that possibly contribute to the excess vibrational modes related to the Boson peaks. Despite the importance of Boson peaks for understanding glass properties, there is no report of Boson peaks in amorphous phase-change materials (PCMs) (e.g. Ge-Sb-Te), probably because their poor glass forming abilities, typically requiring a critical cooling rate of $10^9$ K s$^{-1}$ to obtain amorphous (glassy) phase, hinder the synthesis of bulk amorphous samples for characterizing the Boson peaks.

Amorphous PCMs belong to a family of functional materials for novel devices and photonic applications. Due to their fast crystallization kinetics, PCMs can be rapidly and reversibly switched between amorphous and crystalline phases within tens of nanoseconds. With the large optical and electric property contrast between the two phases, they can be used to encode data for non-volatile memory, photonic, and neuromorphic computing devices[25,26]. Due to their poor glass forming ability, many PCMs (e.g. $Ge_2Sb_2Te_5$, GeTe, and $Ge_{15}Sb_{85}$) do not even exhibit a glass transition before crystallization sets in at a conventional heating rate (~20 K min$^{-1}$) in the differential scanning calorimetry (DSC)[27–31]. Thus, this makes it difficult to characterize the glass transition and supercooled liquids in PCMs, as often done for conventional glasses. As mentioned above, Boson peaks, as a feature observed in various types of glasses, may reflect some essential properties of glass structures and dynamics. However, to date, it is unknown whether Boson peaks exist in amorphous PCMs, and if they do, what are their relevance to the properties and structures of PCMs.

In this work, we explore new compositions $Ge_{15}Te_{85-x}Sb_x$ ($x = 0$ to 85) on the tie line between $Ge_{15}Te_{85}$ and $Ge_{15}Sb_{85}$ on the ternary phase diagram of Ge-Sb-Te (see Fig. 1) across the Yamada line (GeTe-$Sb_2Te_3$)[32]. While $Ge_{15}Te_{85}$ is a good glass former unsuitable for PCM applications[33], $Ge_{15}Sb_{85}$ is a typical Sb-rich PCM with poor glass forming ability[34]. The thermal properties of as-deposited samples are investigated using differential scanning calorimetry to characterize the glass transition, enthalpy relaxation, crystallization, and melting temperatures. By cold pressing of amorphous powder, we can synthesize bulk pellets, bypassing the constraints of sample geometry, for low-temperature heat capacity measurements. The Boson peaks are identified in heat capacity in all compositions and discussed in the context of dynamic defects in the amorphous solids. The characteristics of the Boson peaks at low temperatures can then be related to the enthalpy relaxation and kinetic fragilities at higher temperatures near the glass transitions. The study of the Boson peaks in PCMs may provide new perspectives to understand PCMs.

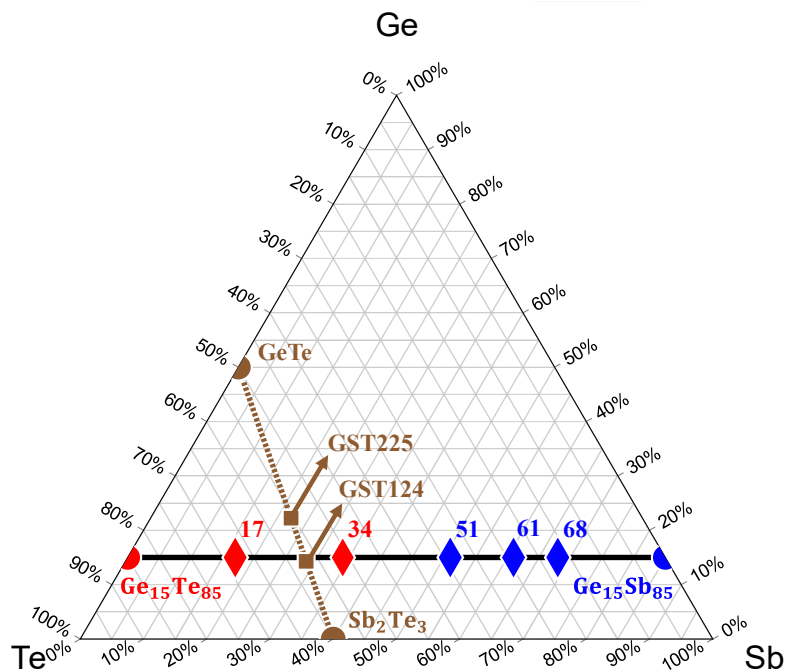

**Figure 1:** Ternary phase diagram of Ge-Te-Sb. A black solid line indicates the pseudo-binary tie line explored by this work between $Ge_{15}Te_{85}$ and $Ge_{15}Sb_{85}$. Each composition of $Ge_{15}Te_{85-x}Sb_x$ is denoted by the respective value of $x$ adjacent to the diamond symbols. Diamonds colored red represent compositions where x < 50, while those colored blue correspond to Sb-rich alloys (x > 50). The brown dotted line indicates the Yamada line between GeTe and $Sb_2Te_3$, associated with typical PCMs $Ge_1Sb_2Te_4$ (GST124) and $Ge_2Sb_2Te_5$ (GST225).

**Table 1:** The compositions and the predominant crystallized phases of $Ge_{15}Te_{85-x}Sb_x$. Crystal phases enclosed in square brackets are inferred by examining the Ge-Te-Sb ternary phase diagram of Bordas et al[35] and comparing the lowest temperature ternary isopleths with relevant binary phase diagrams[35]. The crystalline phases [Sb], [$Sb_2Te_3$], [$\gamma$-SbTe], [$\delta$-$Sb_2Te$], [$\alpha$-GeTe], [GST225], [GST124], and [Te] exhibit a rhombohedral structure at room temperature and ambient pressure, whereas [Ge] has a cubic diamond structure[34,36,37].

| $x$ | Composition | Predominant crystalline phases[35] |
|---|---|---|
| 0 | $Ge_{15}Te_{85}$ | [α-GeTe], [Te] |
| 17 | $Ge_{15}Te_{68}Sb_{17}$ | [GST124], [GST225], [Te] |
| 34 | $Ge_{15}Te_{51}Sb_{34}$ | [γ-SbTe], [$Sb_2Te_3$], [Ge] |
| 51 | Ge$_{15}$Te$_{34}$Sb$_{51}$ | [γ-SbTe], [δ-Sb$_2$Te], [Ge] |
| 61 | Ge$_{15}$Te$_{24}$Sb$_{61}$ | [δ-Sb$_2$Te], [Ge] |
| 68 | Ge$_{15}$Te$_{17}$Sb$_{68}$ | [δ-Sb$_2$Te], [Ge], [Sb] |
| 85 | Ge$_{15}$Sb$_{85}$ | [Ge], [Sb] |

## 2. Methods

### 2.1. Materials synthesis

Each composition in the pseudo-binary tie-line was prepared through magnetron sputtering of in-house synthesized targets. Precursor elements were mixed into quartz ampules in a glovebox with an argon atmosphere. The quartz ampules were sealed with flame under vacuum. The sealed ampules were then placed in a rotating furnace at 50 K above the melting temperature of the corresponding sample for two days for alloying. The compositions of the resulting alloys were determined using X-ray fluorescence (XRF) with a Rigaku NEX CG. The crystalline phases were identified with X-ray diffractions (XRD) using a Rigaku Smartlab diffractometer with a monochromatic Cu Kα X-ray source. Sputtering targets with a one-inch diameter and thickness of ~3 mm were prepared by the Spark-Plasma-Sintering technique (SPS) on the alloy powder. The XRF and XRD analyses were conducted on both surfaces of the synthesized targets and on the powder to verify that the compositions were consistent with the nominal compositions within <1-2 at.%. The substrates for magnetron sputter depositions were 4-inch Si (100) wafers spin-coated with AZ1518 photoresist. The magnetron sputtering deposition system consists of a 1-inch sputtering gun in a vacuum chamber. The sputtering process uses DC 10 W, 10 sccm Ar flow, 5×10$^{-3}$ mbar Ar pressure, and a sputtering time enough for an averaged 3-4 μm film thickness. Amorphous samples were retrieved through lift-off of the film using acetone (wet etching), and the absence of crystals was confirmed with XRD measurements. Further elaboration of the experimental methods is described in the supplementary information.

### 2.1. Characterizations

Differential scanning calorimetry (DSC) studies were performed using a TA Instrument DSC25. The as-deposited samples were crushed into flakes or powder and loaded in an aluminum pan with a sample mass of 3-5 mg. The DSC scans were carried out at a standard heating rate of 20 K min$^{-1}$ from 313 K to 723 K or just below the melting point. The initial scan of the amorphous sample was followed by a rescan of the crystallized sample to obtain the reference scan of the crystalline phase. To determine the melting temperatures $T_m$ and the liquidus temperatures $T_L$, a NETZSCH STA 449 *F3 Jupiter* with alumina pans were used to access the temperature range above 873 K at the same heating rate.

The low-temperature heat capacity was measured with a Quantum Design Physical Property Measurement System (PPMS) in a temperature range from 1.8 K to 300 K. The amorphous powder samples were pressed (<270 MPa) into the pellets with a diameter of 3 mm and a mass of <10 mg. The background signal from the sample platform and thermal paste are subtracted from the measured heat capacity. Calculations and fits were carried out using a customized Python script.

## 3. Results

### 3.1. Enthalpy relaxation, glass transition, and crystallization

Figure 2A shows the results of calorimetric measurements for the as-deposited amorphous samples of $Ge_{15}Te_{85-x}Sb_x$ ($x$ = 0 to 85). The measured heat flow (mW) is normalized to the sample mass and converted to molar heat capacity $C_p$ (J mol$^{-1}$ K$^{-1}$) by multiplying the heating rate (K s$^{-1}$) and the averaged molar mass. The excess heat capacity $C_p^{exc}$ of the amorphous phase over the crystalline was obtained by subtracting the re-scan of the crystallized sample from the initial scan of the as-deposited sample with the same heating rate of 20 K min$^{-1}$ (0.33 K s$^{-1}$). All compositions show one or two sharp exothermic peaks (Fig. S2A for details), indicating the crystallization upon heating. The two crystallization peaks of $Ge_{15}Sb_{85}$ at 512 K and 609 K were attributed to the crystallization of [Sb] and [Ge] phases, respectively[34,38]. The crystallization peaks of the ternary compositions with $x$ > 50 can be attributed to a cubic Ge phase with a diamond structure and a rhombohedral Sb-rich phase ([Sb] or [δ-$Sb_2Te$])[34,39] (see Table 1). Those peaks for alloys with $x$ < 50 correspond to crystalline phases mainly related to those on the Yamada line (i.e. GeTe, GST124, GST225, $Sb_2Te_3$, as indicated in Table 1).

Figure 2B shows the zoom-in view of the $C_p^{exc}$ curves, revealing the thermal events in the glassy states before the crystallization (the full crystallization peaks are not shown here for clarity). Clear endothermic $C_p^{exc}$ jumps are observed in $Ge_{15}Te_{85-x}Sb_x$ ($x$ = 0, 17, 51, 61, and 68), indicating the presence of glass transitions at $T_g$. This implies relatively good amorphous thermal stabilities in these compositions, although the glass transitions might not have fully completed before being cut off by crystallization peaks. However, for both $x$ = 34 (i.e. $Ge_{15}Te_{51}Sb_{34}$) and $x$ = 85 (i.e. $Ge_{15}Sb_{85}$), no glass transition is observed before crystallization sets in at this heating rate. The lack of a calorimetric glass transition is reminiscent of the behavior of PCMs on the Yamada line (e.g. $Ge_2Sb_2Te_5$ and $Ge_1Sb_2Te_4$) due to poor glass forming abilities[32]. Note that $Ge_{15}Te_{51}Sb_{34}$ is compositionally close to $Ge_1Sb_2Te_4$, as shown in the ternary phase diagram of Fig. 1.

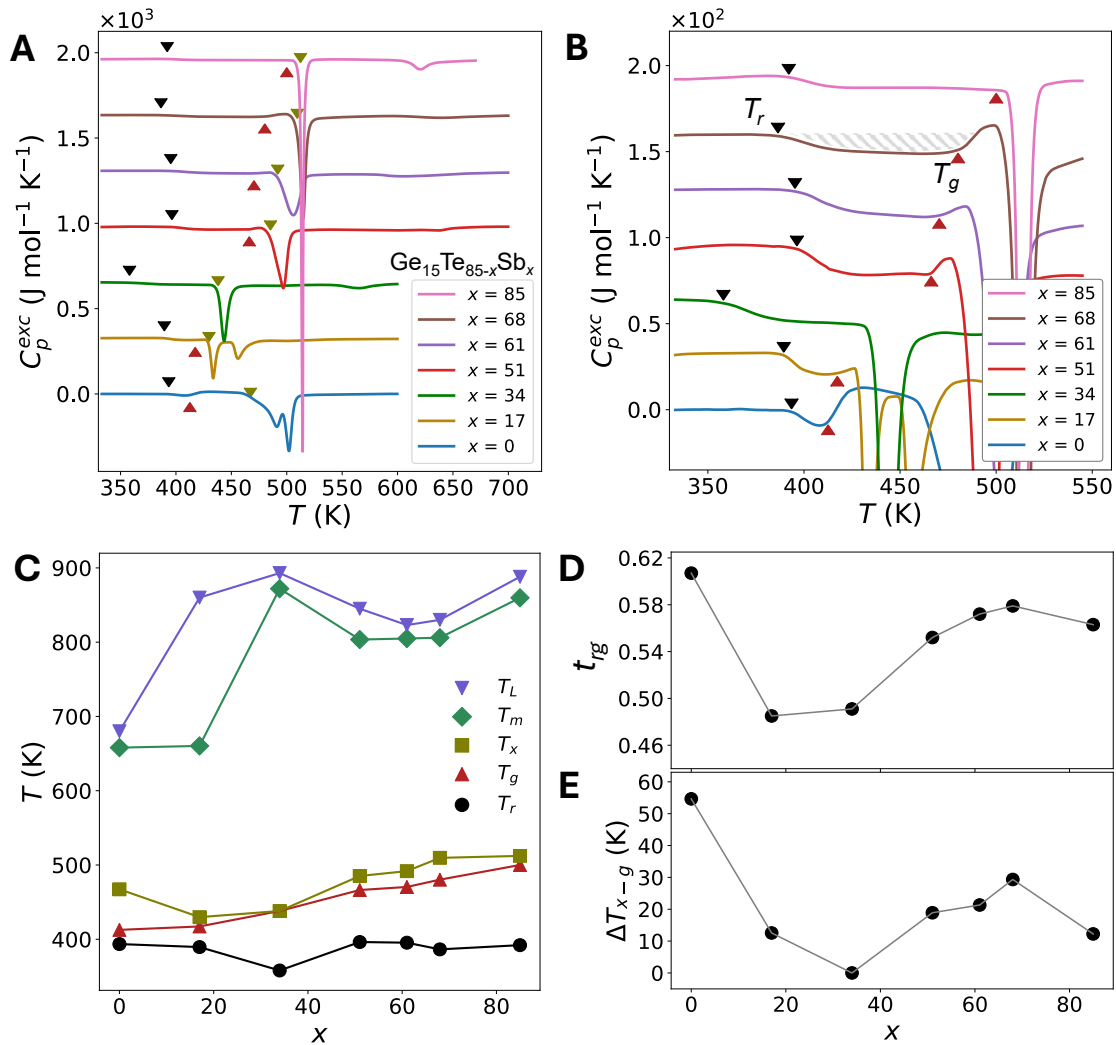

**Figure 2: The differential scanning calorimetric measurements of as-deposited $Ge_{15}Te_{85-x}Sb_x$.** **(A)** Excess heat capacity $C_p^{exc}$ (endothermic up) from DSC scans of as-deposited $Ge_{15}Te_{85-x}Sb_x$ upon heating at 20 K min$^{-1}$ after subtracting the re-scan of crystallized samples. Black downward triangles indicate the onset of enthalpy relaxation ($T_r$); upward red triangles indicate the glass transition temperatures ($T_g$); downward olive triangles represent the onset of crystallization ($T_x$). The curves are vertically shifted by a constant for clarity. **B)** The magnified view of $C_p^{exc}$ highlights the regions of glass transitions and enthalpy relaxations (see the shaded area as an example). The curves are vertically shifted by a constant for clarity. **C)** The characteristic temperatures are plotted against $x$, including $T_r$, $T_g$, $T_x$, the onset of melting ($T_m$), and the liquidus temperature ($T_L$). **D)** The Turnbull parameter $t_{rg} = T_g/T_L$. **E)** The width of the supercooled liquid region, $\Delta T_{x-g} = T_x - T_g$.

What's more, a pronounced exothermic dip is observed below $T_g$ in all compositions, indicative of the enthalpy relaxation in the glassy states. This is because a certain amount of enthalpy that was trapped in the glassy state during vitrification is released upon heating. It is well known that the enthalpy relaxation depends on not only the intrinsic material properties, but also the vitrification processes such as cooling rates[2,3]. The samples of this work were prepared using magnetron sputtering, which can be assigned to have an effective cooling rate proportional to the applied voltage[40]. The voltage was kept at a constant 370(15) V for all compositions. Thus, the enthalpy relaxation should primarily reflect the characteristic behavior of the materials themselves. As shown in Fig. 2B, the width of the sub-$T_g$ exotherms is larger in the Sb-rich alloys ($x > 50$).

We have additionally performed the heat flow measurements for higher temperatures above the melting point using a NETZSCH STA (see Fig. S2B). The onset of the melting $T_m$ and the liquidus temperature $T_L$ are determined and plotted in Fig. 2C together with the onset of crystallization $T_x$, $T_g$, and the onset of enthalpy relaxation $T_r$ obtained from the TA DSC25. The $T_m$ and $T_L$ show a maximum at around $x \approx 34$ and a local minimum around $x \approx 61$. This result is consistent with the ternary phase diagram of Ge-Te-Sb[35].

Figure 2D shows the Turnbull parameter $t_{rg} = T_g/T_L$, which is an effective indicator for the glass forming ability, proposed by Turnbull[41], and widely used for metallic glasses and oxide glasses[19,31]. According to Turnbull, $t_{rg} = 2/3$ indicates a good glass former, whereas $t_{rg} = 0.5$ corresponds to poor glass forming ability[41]. For Ge$_{15}$Te$_{85-x}$Sb$_x$, $t_{rg}$ falls to ~ 0.5 for $x = 17$ to 34. Note that $x \approx 34$ corresponds to the saddle point in the vicinity of the Yamada line (see Fig. 1). $t_{rg}$ increases to the maximum 0.58 for $x = 61$ to 68, indicating a better glass forming ability as expected from the local minimum $T_L$.

Figure 2E shows the width of the supercooled liquid region, $\Delta T_{x-g} = T_x - T_g$, which measures the thermal stability of the amorphous phase. A local maximum $\Delta T_{x-g} = 24$ K is observed for the composition with $x = 68$ (near the liquidus minimum), indicating an optimal amorphous stability. By comparing Fig. 2D and 2E, we can observe a weak correlation between the Turnbull parameter $t_{rg}$ and $\Delta T_{x-g}$. This is not surprising as earlier studies showed that amorphous thermal stability is not necessarily correlated with glass forming ability[42]. The former concerns the propensity of crystallization on heating from glasses (possibly with quenched-in nuclei), whereas the latter is related to the crystallization during cooling from the melt[43].

**Table 2:** The temperatures of the thermodynamic events, including $T_r$, $T_g$, $T_x$, $T_m$, $T_L$, $\Delta T_{g-r}$, $\Delta T_{x-g}$, and $t_{rg}$. Note: *The calorimetric $T_g$ is not present in the DSC scans for $x = 85$; therefore $T_g$ is taken as the α-relaxation peak temperature in loss modulus from a dynamic mechanical spectroscopy study[44]. †Taken from Ref[31]. §Due to the absence of calorimetric $T_g$, an endpoint temperature is chosen just before $T_x$.

| x | $T_r$ (K) | $T_g$ (K) | $T_x$ (K) | $T_m$ (K) | $T_L$ (K) | $\Delta T_{g-r}$ (K) | $\Delta T_{x-g}$ (K) | $t_{rg}$ |
|---|---|---|---|---|---|---|---|---|
| 0 | 394 | 412 | 468 | 658† | 680† | 18 | 56 | 0.607 |
| 17 | 390 | 417 | 430 | 660 | 860 | 27 | 13 | 0.485 |
| 34 | 358 | --- | 438 | 872 | 893 | 80§ | --- | 0.491 |
| 51 | 396 | 467 | 485 | 804 | 846 | 70 | 18 | 0.552 |
| 61 | 396 | 471 | 392 | 806 | 824 | 75 | 21 | 0.572 |
| 68 | 387 | 480 | 510 | 806 | 830 | 93 | 24 | 0.579 |
| 85 | 392 | 500* | 512 | 860 | 889 | 108* | 12* | 0.563 |

## 3.2. The Boson peaks and dynamic defects

The low-temperature heat capacity was measured from 1.8 K to 300 K using PPMS, as described in the method section. Low-temperature solids are commonly described by the Debye model, as an elastic continuum populated primarily by acoustic phonon branches[45]. In the Debye model, the heat capacity is given by the Debye-Einstein equation or the Debye contribution ($C_D$)[45],

$$C_D = 9R\left(\frac{T}{\theta_D}\right)^3 \int_0^{\frac{\theta_D}{T}} \frac{x^4 \cdot e^x}{(e^x-1)^2} dx, \quad (1)$$

where $R$ is the gas constant and $\theta_D$ is the characteristic Debye temperature of the solid. $\theta_D$ represents the highest phonon frequency $\omega_D$ (via $\theta_D = \hbar\omega_D/k_B$, where $k_B$ is the Boltzmann constant) of the ground-state vibrational spectrum of the solid and is related to the average bond stiffness[45]. Thus, at sufficiently low temperatures (<30 K), $C_D$ reduces to $C_D \propto T^3$ as per Debye's 3rd thermodynamic law[45]. If the heat capacity follows the Debye model, it should exhibit a flat line in the plot of $C_p/T^3$ versus $T$. $C_p/T^3$ is also referred to as Debye-reduced heat capacity.

However, as shown in Figure 3A, a pronounced peak, referred as the Boson peak, is observed below 10 K in all compositions, demonstrating an excess contribution over the Debye model.

The Boson peaks suggest the existence of excess vibrational modes in their glassy state, as a universal feature in amorphous solids[1,5]. With increasing Sb-content (i.e., higher $x$ values), the magnitude of the Boson peaks lowers, and the position of the Boson peaks shifts to higher temperatures.

Studies on bulk metallic glasses have commonly modelled the Boson peak in heat capacity by treating the excess vibrational modes originating from the Einstein model[24,46]. It is assumed that a low-temperature amorphous solid behaves like a Debye solid where some pockets, clusters, or regions exhibit low-frequency modes similar to the harmonic oscillators of the Einstein model (i.e., Einstein pockets)[1,3,24,46]. The Boson peak can then be fitted as the sum of the Debye contribution $C_D$ and an excess term.

$$C_p = C_D + a_{exc} \cdot \int_0^\infty C_E \cdot \chi(\theta_E) \, d\theta_E \qquad (2)$$

where $C_E$ is the contribution to heat capacity from the Einstein model[45] and $a_{exc}$ is the coefficient for the excess term as a free fitting parameter, representing the amount of excess vibrational modes accessible by the solid. The Einstein model describes a solid as a collection of independent harmonic oscillators, as temperature approaches the Dulong-Petit limit. The $C_E$ is given by.

$$C_E = 3R \left(\frac{\theta_E}{T}\right)^2 \cdot \frac{e^{\frac{\theta_E}{T}}}{\left(e^{\frac{\theta_E}{T}} - 1\right)^2} \qquad (3)$$

where $\theta_E$ is the characteristic Einstein temperature of the solid, and is the lower temperature limit of Equation 3, corresponding to the lowest excitation energy of the harmonic oscillator modes with frequency $\omega_E$ via $\theta_E = \hbar\omega_E/k_B$. Thus, the heat capacity around $\theta_E$ corresponds to the activation of the harmonic oscillator modes. Their subsequent saturation results in the Dulong-Petit limit[45]. Due to the spatial heterogeneity of amorphous solids, the distribution of $\theta_E$ can be assumed to follow a Gaussian distribution,

$$\chi(\theta_E) = \frac{1}{\sigma_E \cdot \sqrt{2\pi}} \cdot e^{\frac{-(\overline{\theta_E} - \theta_E)^2}{2\sigma_E^2}} \qquad (4)$$

where $\chi(\theta_E)$ is the Gaussian probability density function of $\theta_E$ with the mean $\overline{\theta_E}$ and the standard deviation $\sigma_E$. The integral of the product between Equation 3 and 4, results in the expectation value of an Einstein contribution with a Gaussian distribution of Einstein temperatures[47].

Thus, Equation 2 contains four fitting parameters ($\theta_D$, $a_{exc}$, $\overline{\theta_E}$, $\sigma_E$), where each of their effect on the Boson peak shape is illustrated in Fig. S3. Note that Equation 2 slightly differs from those used in the literature[24,46], where the $C_p$ would be expressed as the weighted sum of the Debye and Einstein contributions[3,24,46]. Here the second term is considered as an additional contribution to the Debye contribution. This implies that instead of viewing the two types of regions (i.e. Einstein pockets and the Debye solid) as being completely separated, the phonons, described by the Debye model, propagate throughout the whole solid, where some local regions would exhibit the low-frequency Einstein modes as the phonons pass through them. These local regions might be regarded as string-like dynamic defects with quasi-localized vibrations[1,9], or loosely bound atoms (or clusters[48]) in flow units[24,49].

The experimental data are fitted with Equation 2, where the experimental data and fitted result are plotted in Fig. 3A, with the resulting fitting parameters displayed in Fig. 3B-E and tabulated in Table 3.

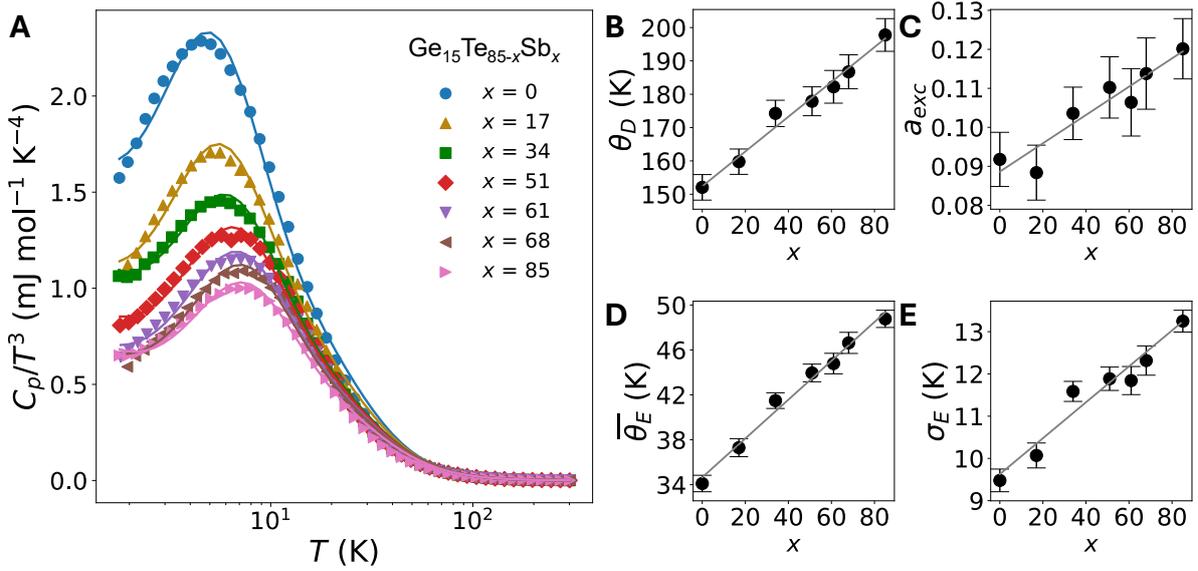

**Figure 3: The low-temperature heat capacity measurements of as-deposited Ge$_{15}$Te$_{85-x}$Sb$_x$.** **(A)** The Boson peaks in Debye-reduced heat capacity, $C_p/T^3$. The solid lines represent the data fits with Equation 2. **(B)** The resulting Debye temperature $\theta_D$. **(C)** The coefficient of excess modes related to the Einstein contribution $a_{exc}$. **(D)** The mean Einstein temperature $\overline{\theta_E}$ and **(E)** its standard deviation $\sigma_E$. All parameters exhibit a linear relationship (grey lines) with the substitution variable $x$, for which the linear fits yield $\theta_D = 0.52(6) x + 152(3)$, $a_{exc} = 0.4(1) \times 10^{-3} x + 0.09(5)$, $\overline{\theta_E} = 0.17(1) x + 34.7(5)$, and $\sigma_E = 0.043(4) x + 9.6(2)$.

The $\theta_D$, $a_{exc}$, $\overline{\theta_E}$ and $\sigma_E$ all increase linearly with increasing Sb-content $x$. The linear relationships suggest that the low-temperature vibrational properties are not directly correlated with the glass forming ability and thermal stabilities which exhibit nonmonotonic behaviors with $x$ as shown in Fig. 2D and 2E. Both $\theta_D$ and $\overline{\theta_E}$ are proportional to the characteristic vibrational frequencies $\omega_D$ and $\omega_E$ related to the overall solid and the Einstein pockets, respectively. Thus, the increase in $\theta_D$ and $\overline{\theta_E}$ is probably due to the substitution of Te atoms with the lower mass Sb atoms involved in the vibrations. The increase of $a_{exc}$ with $x$ indicates a larger contribution of the Einstein modes possibly resulting from a larger region of Einstein pockets or dynamic defects distributed in the solid. This implies a higher structural or dynamic heterogeneity in the amorphous alloys with a higher Sb-content. Meanwhile, a large $\sigma_E$ suggests a broader distribution of vibrational frequencies of the Einstein modes. Interestingly, the coefficient of variation ($\sigma_E/\overline{\theta_E}$) retains approximately a constant value, 27.1(5) %, despite the varying $\overline{\theta_E}$.

**Table 3:** The results of fits to the Boson peak using Equation 2: the Boson peak temperature ($T_{Bp}$), the maximum Debye-reduced heat capacity $C_p/T_{Bp}^3$, the Debye temperature ($\theta_D$), the excess coefficient ($a_{exc}$), the mean Einstein temperature ($\overline{\theta_E}$), and the standard deviation of the Einstein temperature ($\sigma_E$). The parentheses next to the fit parameters refer to the error bars of the fit at the last digit, e.g., 152(4) = 152 ± 4 and 9.5(3) = 9.5 ± 0.3.

| $x$ | $T_{Bp}$ (K) | $C_p/T_{Bp}^3$ (mJ mol$^{-1}$ K$^{-4}$) | $\theta_D$ (K) | $a_{exc}$ | $\overline{\theta_E}$ (K) | $\sigma_E$ (K) |
|---|---|---|---|---|---|---|
| 0 | 4.53 | 2.29 | 152(4) | 0.092(7) | 34.1(7) | 9.5(3) |
| 17 | 5.18 | 1.71 | 159(4) | 0.088(7) | 37.3(8) | 10.1(3) |
| 34 | 5.59 | 1.46 | 174(4) | 0.104(7) | 41.5(7) | 11.6(2) |
| 51 | 6.37 | 1.28 | 178(4) | 0.110(8) | 44.0(8) | 11.9(3) |
| 61 | 7.09 | 1.15 | 182(5) | 0.106(9) | 44.8(9) | 11.8(3) |
| 68 | 7.15 | 1.04 | 187(5) | 0.114(9) | 46.6(9) | 12.3(3) |
| 85 | 7.12 | 1.00 | 198(5) | 0.120(8) | 48.8(8) | 13.2(3) |

## 4. Discussion

### 4.1. Connecting the low-temperature Boson peaks to the higher-temperature thermodynamics.

Boson peaks have been studied in various types of glasses including, oxides[50–53], molecular glasses[2,54], polymers[55], bulk metallic glasses [3,13,24,46,56,57], and here PCMs. Recent studies of 2- and 3-dimensional model systems revealed that the Boson peaks, the fast β- and α-relaxations all share a common structural origin as string-like dynamic defects[1,9]. As the temperature rises, vibrations are dampened due to anharmonicity, and structural relaxations emerge as the dominant dynamic process[1]. Thus, it is plausible to anticipate a correlation between the Boson peaks and relaxation behaviors. Figure 4A shows the width of the sub-$T_g$ enthalpy relaxation $\Delta T_{g-r} = T_g - T_r$ plotted against $a_{exc}$. A larger $\Delta T_{g-r}$ is correlated with a larger $a_{exc}$. This implies that the substituting Te with Sb atoms leads to a larger fraction of string-like dynamic defects, which underlie the increase in both $a_{exc}$ and $\Delta T_{g-r}$.

To understand the Boson peaks in a broad context, we have analyzed various types of glasses whose Boson peaks are available in the literature using the same method as described above (i.e. Equation 2). As shown in Fig. 4C-F, these glasses include PCMs, oxides[50–53], and metallic glasses (see Table S1 for all fitting parameters). The metallic glasses consist of CuZr-based[3,56], PdNiP-based[3], Zr-based[3,57], La-based[46] alloys, and one high entropy bulk metallic glass [13]. The resulting fitting parameters are investigated with respect to their $T_g$ (Fig. 4B-C).

The Debye temperature $\theta_D$ of crystalline materials has been related to the melting temperature $T_m$, according to Lindemann's melting criteria, $T_m = Mk_B\theta_D^2 <\mu_{cri}^2>/9h^2$, where $M$ is the atomic mass, $k_B$ is the Boltzmann constant, $\mu_{cri}$ is the critical mean-square thermal displacement, and $h$ is the Planck constant[58]. Since both melting and glass transition involve a solid-to-liquid transition, previous studies proposed a relation between $\theta_D$ and $T_g$ in metallic glasses with the Lindemann-type formalism, $T_g = aM\theta_D^2 + b$, where $a$ and $b$ are constant[58]. It was found that $T_g$ is roughly proportional to $M\theta_D^2$ in many metallic glasses[59]. This equation assumed that the $\mu_{cri}$ at $T_g$ and $T_m$ are the same, although the assumption is hardly accurate. In Fig. 4B, we plot $M\theta_D^2$ versus $T_g$ for PCMs, where a linear relation is evident with $a = 5.5(6) \times 10^{-5}$ and $b = 256(21)$. The similar analysis is applied to metallic and oxide glasses and the results are shown with PCMs in Fig. 4C, where two linear fits are performed for the oxides and the rest, respectively, yielding $a = 2.1(7) \times 10^{-4}$ and $b =$ 306(259) for oxides (red symbols); $a = 6.4(8) \times 10^{-5}$ and $b = 206(44)$ for the rest. The PCMs appear to fall within the same group as metallic glasses; however, they are notably different from oxide glasses despite both being covalently bonded.

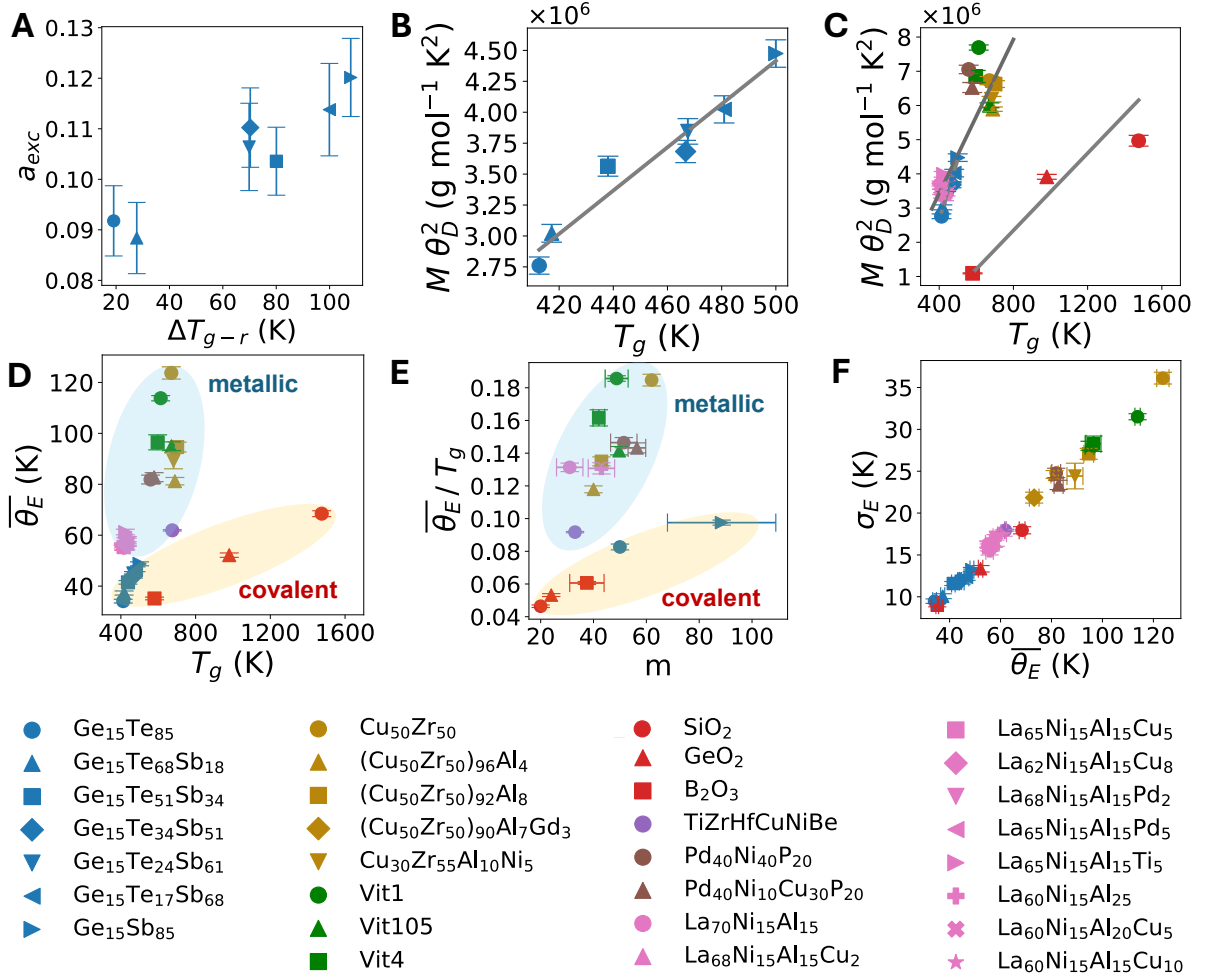

**Figure 4:** **The Boson peak analysis using Equation 2 on a variety of covalent and metallic glasses.** A) Coefficient of excess modes ($a_{exc}$) plotted against the width of the enthalpy relaxation ($\Delta T_{g-r}$) for $Ge_{15}Te_{85-x}Sb_x$. B) The relationship between $M\theta_D^2$ and $T_g$ for $Ge_{15}Te_{85-x}Sb_x$. C) The relationship between $M\theta_D^2$ and $T_g$ for PCMs, oxide, and metallic glasses. The $T_g$ of metallic and oxide glasses are taken from references listed in Table S1. D) The mean Einstein temperature $\overline{\theta_E}$ against $T_g$, where metallic and covalent bonding types are marked by the light blue and orange areas, respectively. E) $\overline{\theta_E} / T_g$ against the fragility index ($m$) for metallic glasses (light blue area) and covalent glasses (orange area). The references for $m$ are given in Table S1. F) The standard deviation of the Einstein temperature ($\sigma_E$) plotted against the mean Einstein temperature ($\overline{\theta_E}$) for all analyzed Boson peaks.

Figure 4D shows the relationship between $\overline{\theta_E}$ and $T_g$. While $\overline{\theta_E}$ is the mean activation temperature of the low-frequency vibrations of the Einstein pockets, $T_g$ corresponds to the activation temperature of α-relaxations[44] and plastic flow[22]. The data point cluster of metallic glasses separates from that of oxide glasses. PCMs appear to lie in between. This implies that the underlying relation between the excess vibrational modes and relaxations depends on the bonding types of glasses. Most metallic glasses have a high $\overline{\theta_E}$ and a moderate $T_g$, whereas the covalently bonded oxide glasses may have a high $T_g$ with a low $\overline{\theta_E}$. Although PCMs are usually also considered being covalently bonded in their glassy states, both of their $\overline{\theta_E}$ and $T_g$ are low. This is reminiscent of the difference in slow β-relaxation behaviors between PCMs and non-PCM covalent glasses, as discussed in studies[44,60]. These studies suggested that the bonding character of PCMs is less covalent than that of conventional covalent rigid network glasses[44,60]. Within the same bonding type, $\overline{\theta_E}$ seems to positively correlate with $T_g$.

In Fig. 4E, we take the ratio of $\overline{\theta_E}$ to $T_g$ as a unitless parameter, and plot it against the fragility index $m$. A correlation between $\overline{\theta_E}/T_g$ and $m$ can be identified, which also depends on the bonding type of amorphous solids. The slope of $\overline{\theta_E}/T_g$ versus $m$ for PCMs and oxide glasses appears shallower than that for metallic glasses. We find that a higher $\overline{\theta_E}$ always corresponds to a larger standard deviation $\sigma_E$ of $\overline{\theta_E}$, whose ratio $\sigma_E/\overline{\theta_E}$ keeps a universal constant of 28(1)% (Fig. 4F) for all the analyzed Boson peaks across different glass types. This implies that the system with a higher $\overline{\theta_E}$ has a broader distribution of the vibrational frequencies of Einstein pockets and thus more heterogenous vibrational dynamics. In this context, one possible interpretation of the correlation in Fig. 4E is that the heterogeneity in vibrational dynamics at low temperatures is transformed into the heterogeneity in relaxation dynamics at elevated temperatures. The latter is known to be associated with a higher kinetic fragility[16]. More data points of fragility for PCMs are desirable to consolidate the apparent linear correlation between $\overline{\theta_E}/T_g$ and $m$. This might provide a way to estimate the fragility of PCMs, which is difficult to characterize in their supercooled liquid state due to the interference of fast crystallization. We note that earlier studies showed that the height of Boson peaks exhibit a rough decreasing trend with increasing fragility in various metallic and non-metallic glasses[5,10]. Their approaches took the maximum value of the Debye-reduced heat capacity, $C_p/C_D$, without modelling the Boson peaks. It was also observed in this work (Fig. 3A) and others involving elemental substitutions[46,50] that a higher Boson peak is correlated with a lower peak temperature. Yet, when considering a wider range of systems [3,13,24,46,50–53,56,57], the Boson peaks could display different heights but with the same peak temperature. As shown in Fig. S3, the Boson peak height is influenced by several parameters i.e., $\theta_D$, $a_{exc}$, $\overline{\theta_E}$ and $\sigma_E$. Comparing the Boson peak height of different glass types is not necessarily equivalent to comparing the same parameter. Thus, a fitting to the Boson peaks using Equation 2 allows for extracting individual parameters of the excess vibrational modes and correlating them with other properties.

## 5. Conclusion

Low-temperature heat capacity measurements reveal the presence of a Boson peak below 10 K in the pseudo-binary PCM compositions along the tie-line between $Ge_{15}Te_{85}$ and $Ge_{15}Sb_{85}$ within the Ge-Sb-Te ternary phase diagram. The features of the Boson peaks systematically vary with a systematic substitution of Te by Sb atoms, despite the nonmonotonic variations in glass-forming ability and thermal stability. These Boson peaks can be interpreted through the concept of dynamic defects, linking the low-temperature Boson peaks to structural relaxation behaviors observed at higher temperatures. Metallic and conventional covalent glasses display distinct characteristic behaviors of Boson peaks. Their correlations with $T_g$ and $m$ depend on bonding types. Despite amorphous PCMs commonly being regarded as covalently bonded, they appear, as "metalloid glasses", to exhibit properties intermediate to those of metallic and conventional covalent glasses, or sometimes align even with metallic glasses. In their crystalline counterpart, the property portfolio of PCMs' crystalline phases has been shown to differ from metallic and covalent compounds[61]. Future studies would benefit from characterizing the Boson peaks in those typical PCM compositions on the Yamada line, including GeTe, $Ge_1Sb_2Te_4$, and $Ge_2Sb_2Te_5$. Thermal treatments such as annealing, which modify the glassy state of materials, should be examined for their effects on Boson peaks. Inelastic neutron scattering experiments may allow for directly determining the excess vibrational density of states of PCMs. An earlier study reported a weak anomalous hump at low-temperature heat capacity even in crystalline elemental antimony[62], although its origin is not fully understood. Thus, it is also worth investigating the low-temperature heat capacity in the crystalline phases of PCMs. Further studies of vibrational dynamics in both amorphous and crystalline phases might provide insight into bonding characteristics and anharmonicity of PCMs.

**Acknowledgements**

The authors are supported by the research grant (42116) from VILLUM FONDEN. The project is partially supported by Novo Nordisk Fonde (NNF21OC0071257). The authors declare no conflict of interest. Author contributions: SW conceived the project. JM performed experiments and analyzed data with input from SW and TF. JM and SW wrote the manuscript with input and editing from TF.

# Supplementary information of "Unveiling the Boson Peaks in Amorphous Phase-Change Materials"

Jens Moesgaard[1], Tomoki Fujita[1], and Shuai Wei[1,2]

[1]Department of Chemistry, Aarhus University, 8000 Aarhus, Denmark.

[2]iMAT Centre for Integrated Materials Research, Aarhus University, 8000 Aarhus, Denmark.

## 1. Powder X-ray diffraction measurement

We performed powder X-ray diffraction experiment for the as-deposited samples to confirm the amorphous state. The samples were mounted on the sample stage of Rigaku Smartlab diffractometer with Co $K_\alpha$ X-ray source, a 2 mm incident slit, and a lead knife edge. Figure S1 shows the measured diffraction patterns with the diffuse scattering peaks around 2.0 Å$^{-1}$, which are characteristic for amorphous structures.

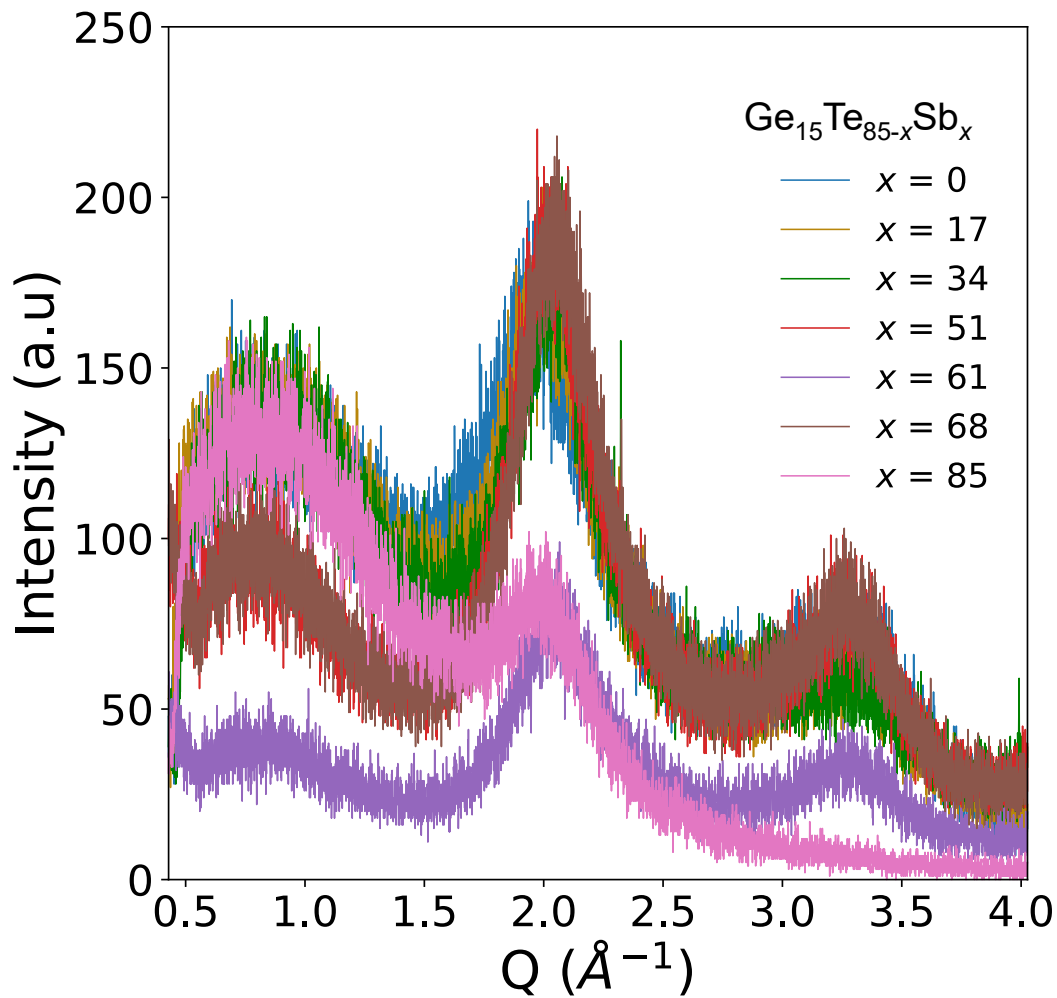

**Figure S1:** Powder x-ray diffraction data of the as-deposited samples.

## 2. Pressing of pellets for Boson peak measurements

To measure the low-temperature Boson Peak on the Physical Property Measurement System (PPMS), the amorphous sample powders were compressed into pellets with a diameter of 3 mm and weight of <10 mg. The pressure die set has a diameter of 3 mm with a max yield of 270 MPa, where pressure was applied manually to avoid possible pressure-induced crystallization.

## 3. The DSC scans at higher temperatures and above the melting point

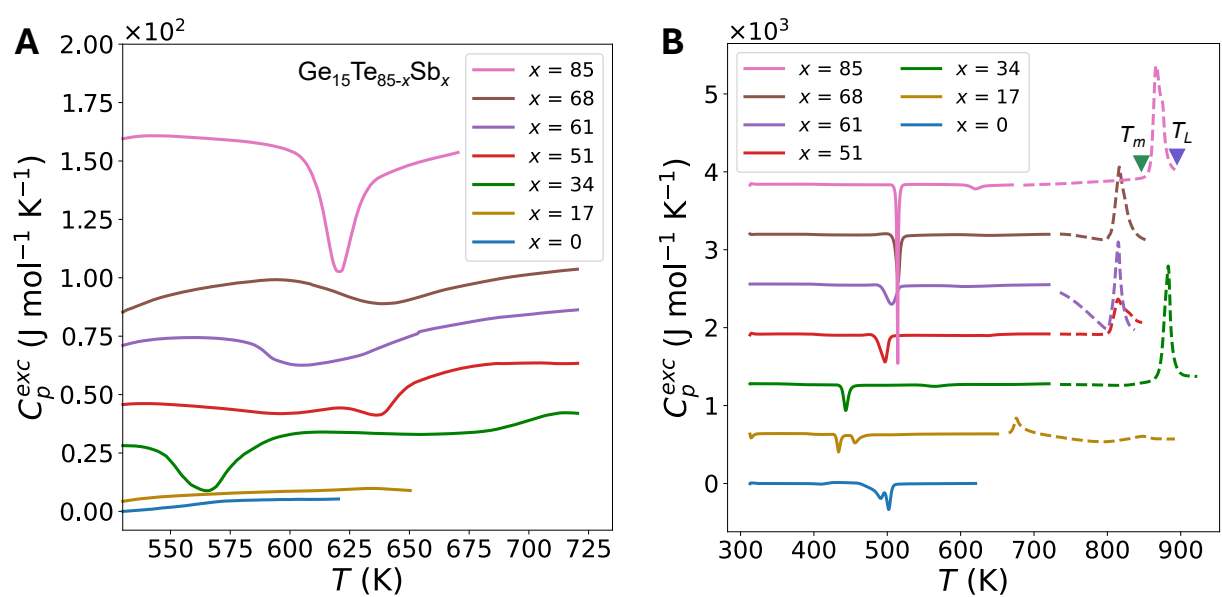

**Figure S2:** Excess heat capacity (endothermic up) from DSC scans at 20 K min$^{-1}$ after subtracting the rescan of crystallized samples. **A)** The temperature range from 550 to 725 K is shown here to magnify the thermal event of the small secondary crystallization peaks above the temperature of the first primary crystallization peaks. **B)** Full temperature range of excess heat capacity, combining the scans at high temperature (dashed lines) using Netzsch STA/DSC to determine $T_m$ and $T_L$. For $x = 17$, the onset of melting at ~660 K is well separated from the completion of melting at ~860 K, which is consistent with the reported phase diagram (Fig. 6 of [1]). Curves are vertically shifted for clarity.

## 4. Analysis of Boson peaks.

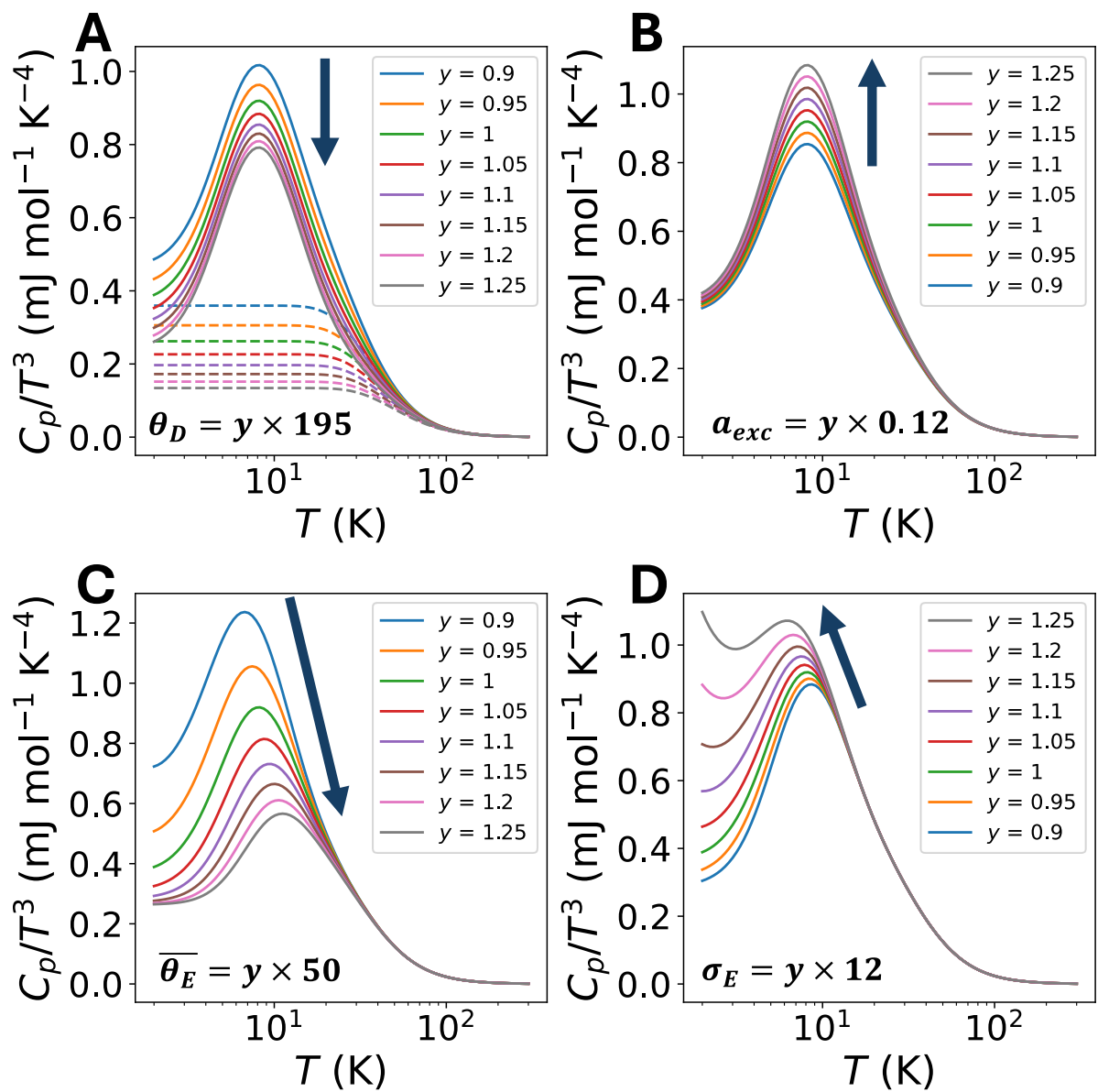

**Figure S3:** The simulation of the Boson peak changes influenced by individual parameters of Equation 2 (see main text). Each parameter varies by a factor of *y* between 0.9 and 1.25. An arrow indicates the direction of increasing *y*. The dashed lines in **A)** represent the corresponding Debye contribution.

**Table S1:** The resulting fitting parameters of Equation 2 (see main text) for the Boson peaks of all analyzed glasses, the Debye temperature ($\theta_D$), the excess coefficient ($a_{exc}$), the mean Einstein temperature ($\overline{\theta_E}$), and the standard deviation of the Einstein temperature ($\sigma_E$), as well as the tabulated glass transition temperature ($T_g$), and fragility index ($m$). If a $T_g$ or $m$ is not cited, then its reference belongs to the one in the "Glass" column. The parentheses next to any variable refer to the error bars of the fit at the last digit, or in the case of $T_g$ and $m$ the uncertainty after averaging multiple values found in the literature. *Debye temperature was too high to fit, as the effect of the Debye temperature on Debye-reduced heat capacity becomes negligible for Debye temperatures above 800 K.

| Glass | $\theta_D$ (K) | $a_{exc}$ | $\overline{\theta_E}$ (K) | $\sigma_E$ (K) | $T_g$ (K) | $m$ |
|---|---|---|---|---|---|---|
| $Ge_{15}Te_{85}$ | 152(4) | 0.092(7) | 34.1(7) | 9.5(3) | 413 | 50[2] |
| $Ge_{15}Te_{17}Sb_{68}$ | 159(4) | 0.088(7) | 37.3(8) | 10.1(3) | 418 | --- |
| $Ge_{15}Te_{51}Sb_{34}$ | 174(4) | 0.104(7) | 41.5(7) | 11.6(2) | 438 | --- |
| $Ge_{15}Te_{34}Sb_{51}$ | 178(4) | 0,110(8) | 44.0(8) | 11.9(3) | 467 | --- |
| $Ge_{15}Te_{24}Sb_{61}$ | 182(5) | 0.106(9) | 44.8(9) | 11.8(3) | 468 | --- |
| $Ge_{15}Te_{17}Sb_{68}$ | 187(5) | 0.114(9) | 46.6(9) | 12.3(3) | 481 | --- |
| $Ge_{15}Sb_{85}$ | 198(5) | 0.120(8) | 48.8(8) | 13.2(3) | 500 | 88(20)[3] |
| $Cu_{50}Zr_{50}$[4] | 294(5) | 0.11(1) | 124(2) | 36.1(7) | 670[5] | 62 |
| $(Cu_{50}Zr_{50})_{96}Al_4$[4] | 279(3) | 0.058(5) | 81(1) | 24.6(5) | 689[5] | 40 |
| $(Cu_{50}Zr_{50})_{92}Al_8$[4] | 300(4) | 0.085(8) | 94(2) | 27.1(7) | 701[5] | 43 |
| $(Cu_{50}Zr_{50})_{90}Al_7Gd_3$[4] | 268(3) | 0.048(5) | 73(2) | 21.84(6) | --- | 30 |
| $Cu_{30}Zr_{55}Al_{10}Ni_5$[6] | 287(4) | 0.061(8) | 89(3) | 24.4(2) | 682 | --- |
| Vit1[7] | 358(3) | 0.11(1) | 114(1) | 31.5(5) | 613[8] | 49(4)[8] |
| Vit105[9] | 291(3) | 0.069 (6) | 95(2) | 28.0(9) | 671[10] | 50[10] |
| Vit4[4] | 339(8) | 0.083(1) | 96(2) | 28.3(4) | 597[8] | 42[8] |
| $SiO_2$[11] | 498(16) | 0.032(2) | 68(1) | 17.9(4) | 1475[12] | 20[4] |
| $GeO_2$[13] | 335(6) | 0.026(2) | 52.2(9) | 13.4(4) | 980[14] | 24[4] |
| $B_2O_3$[15] | 280(4) | 0.011(5) | 35.2(5) | 9.0(2) | 580[16] | 38(7)[4] |
| TiZrHfCuNiBe[17] | ---* | 0.011(1) | 61.9(3) | 18.0(1) | 675 | 33[18] |
| $Pd_{40}Ni_{40}P_{20}$[4] | 312(5) | 0.078(6) | 82(2) | 24.8(6) | 559[8] | 51(5)[8] |
| $Pd_{40}Ni_{10}Cu_{30}P_{20}$[4] | 298(7) | 0.083(8) | 83(2) | 23.3(6) | 578[8] | 56(3)[8] |
| $La_{70}Ni_{15}Al_{15}$[19] | 175(4) | 0.13(1) | 55(1) | 16.3(3) | 420 | 31(5)[20] |
| $La_{68}Ni_{15}Al_{15}Cu_2$[19] | 179(4) | 0.13(1) | 55(1) | 16.2(4) | 416 | --- |
| $La_{65}Ni_{15}Al_{15}Cu_5$[19] | 185(5) | 0.12(1) | 55(1) | 15.9(4) | 415 | --- |
| $La_{62}Ni_{15}Al_{15}Cu_8$[19] | 189(4) | 0.12(1) | 56(1) | 15.8(3) | 413 | --- |
| $La_{68}Ni_{15}Al_{15}Pd_2$[19] | 183(4) | 0.16(1) | 58(1) | 17.1(3) | 426 | --- |
| $La_{65}Ni_{15}Al_{15}Pd_5$[19] | 182(4) | 0.14(1) | 57(1) | 16.44(3) | 434 | --- |
| La$_{65}$Ni$_{15}$Al$_{15}$Ti$_5$[19] | 194(3) | 0.18(1) | 61(1) | 17.6(4) | 424 | --- |
| La$_{65}$Ni$_{15}$Al$_{25}$[19] | 183(5) | 0.11(1) | 56(1) | 15.5(5) | 426 | 43(5)[20] |
| La$_{60}$Ni$_{15}$Al$_{20}$Cu$_5$[19] | 185(4) | 0.11(1) | 57(2) | 16.4(5) | 434 | --- |
| La$_{60}$Ni$_{15}$Al$_{15}$Cu$_{10}$[19] | 190(5) | 0.12(1) | 58(1) | 16.5(5) | 434 | --- |